\documentclass[twocolumn,showpacs,preprintnumbers,amsmath,amssymb,prb]{revtex4-1}

\usepackage[dvips,colorlinks=true,bookmarks=false,citecolor=blue,urlcolor=blue]{hyperref} %latex w/dvips

\usepackage{amsmath,amssymb}
\usepackage{graphicx}
\usepackage{verbatim} % for multiple-line comment
\usepackage{color}

\usepackage[english]{babel}
\usepackage{bm}
\usepackage{natbib}
\usepackage{graphicx}

\def\dfrac{\displaystyle\frac}  % Use \displaystyle in the beginning for bigger expressions.

\renewcommand {\phi}{\varphi}

\newcommand{\nix}[1]{}

\usepackage{color}	

\begin{document}
\title{Second-harmonic generation in subwavelength graphene waveguides}
% \author{ }
% \affiliation{}
% \pacs{ }

\author{Daria Smirnova}
\author{Yuri S. Kivshar}

\affiliation{
Nonlinear Physics Center, Australian National University, Canberra ACT 0200, Australia}

% \pacs{73.20.Mf, 42.65.Ky, 78.67.Bf}
% \pacs{42.65.Wi, 78.67.Wj, 73.25.+i, 78.68.+m}

\begin{abstract}
We suggest a novel approach for generating second-harmonic radiation in subwavelength graphene waveguides. We demonstrate that quadratic phase matching between the plasmonic guided modes of different symmetries can be achieved in a planar double-layer geometry when conductivity of one of the layers becomes spatially modulated. We predict theoretically that, owing to graphene nonlocal conductivity, the second-order nonlinear processes can be actualized for interacting plasmonic modes with an effective grating coupler to allow external pumping of the structure and output of the radiation at the double frequency.
\end{abstract}

\maketitle

\section{Introduction}

Graphene plasmonics is the rapidly developing field of nanophotonics that attracted a lot of attention during the past two years~\cite{RevGrigorenko, RevBao, JablanReview, RevLuo, Stauber_Review, Low_Review, Abajo_Review_ACSPhot}. One of the recent novel directions in this field is the study of nonlinear phenomena given the fact that strong confinement of surface p-plasmons in graphene can enhance light-matter interaction~\cite{Abajo176} and facilitate nonlinear response~\cite{PRL_SinglePhoton}. In particular, the recent theoretical studies included the analysis of spatial plasmon solitons in graphene layers~\cite{LPR, DissipSoliton_LPR}, nonlinear difference frequency generation of THz surface plasmons~\cite{PRL_Tokman_2014}, and single-photon operation in a graphene cavity~\cite{PRL_SinglePhoton}.

Double-layer graphene waveguides are known to support symmetric and antisymmetric plasmonic guiding modes classified with respect to the in-plane electric field profile\cite{Hanson, buslaev_JETP, Coupler_PRB}, so we may expect the coupling  between different modes when the nonlinear response becomes important. In this paper, we suggest and study analytically a novel approach for the second-harmonic generation in double-layer graphene waveguides taking the advantage of the possibility to modulate spatially the conductivity of one of the graphene layers by doping or electrostatic gating~\cite{Gating_doping,Engheta,Chemical_doping} thus assisting the coupling of freely propagating light to plasmons in graphene~\cite{bludov_ScatMod_2012, bludov_PhotCr_2012, Nikitin_grating_2013,bludov_primer_2013}.

%%%%%%%%%%%%%%%%%%%%%%%%%

Phased matching between parametrically interacting waves is known to be a crucial requirement for the efficient second-order nonlinear effects~\cite{Boyd}, and in our geometry the phase matching can be achieved by tailoring the modal dispersion of the graphene waveguide enabling plasmon-to-plasmon frequency conversion. More specifically, we demonstrate below
that the phase matching becomes possible between an antisymmetric fundamental frequency (FF) mode and a symmetric second-harmonic (SH) mode in a double-layer graphene waveguide.

\begin{figure}[!b]\centering
  \includegraphics[width=0.95\linewidth]{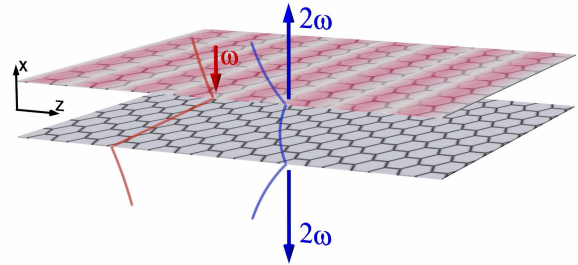} % Fig1_Tamm
  \caption{(Color online) Geometry of the problem. A free-standing waveguide created by two graphene layers is
  illuminated by an external wave being coupled to the phase-matched guided modes of the graphene waveguide. Red curve corresponds to an antisymmetric mode at the fundamental frequency, and a blue curve -- to a second-harmonic symmetric mode (shown with the tangential electric field component). Conductivity of the upper layer is periodically modulated.}
  \label{fig:fig1}
\end{figure}

We employ a theoretical analysis based on the perturbation theory and describe the second-order nonlinear process with graphene plasmons in terms of slowly varying modes of a waveguide structure, the approximation frequently used in the nonlinear optics~\cite{book}. For definiteness, we consider a planar geometry shown schematically in Fig.~\ref{fig:fig1}, where a graphene double-layer waveguide is placed into homogeneous surrounding medium with the dielectric permittivity $\varepsilon$ being illuminated by light from the upper half-space $x>d/2$. As a matter of fact, in our analysis we distinguish two parts of the problem and examine them sequentially,
\begin{itemize}

\item
{\em Linear scattering.} Radiation at the fundamental frequency normally incident onto the structure is scattered by the periodic conductivity grating, and then excites resonantly an antisymmetric mode of the graphene waveguide;

\item
{\em Nonlinear plasmon-to-plasmon conversion.} Due to the second-order nonlocal nonlinearity of graphene~\cite{Glazov2011_SHG, Mikhailov2011_SHG, Glazov2013}, the induced current of the antisymmetric mode serves as a source for the phase-matched symmetric mode at the double frequency which eventually radiates into free space.

\end{itemize}

Similar to the assumptions employed earlier~\cite{DissipSoliton_LPR, Coupler_PRB, Gorbach2013}, namely
low dissipation, weak nonlinearity, and small phase mismatch, here we solve  Maxwell's equations following the procedure of the asymptotic expansion, and demonstrate that the second-order nonlinear processes can be achieved
with conductivity modulation to allow external pumping to couple to the guided modes of the structure generating
output radiation at the doubled frequency.

\section{Linear scattering}

We start our derivation by studying the linear resonant excitation of antisymmetric plasmons,
and write the corresponding system of Maxwell's equations in the form
\begin{equation}
\begin{aligned} \label{eq:eqMaxw2}
&\nabla \times {\bf E} =  - \frac{1}{c} \frac{\partial {\bf H}}{\partial t}\:,\\
&\nabla \times {\bf H} = \frac{\varepsilon }{c} \frac{\partial {\bf E}}{\partial t}\!+\!\frac{4\pi }{c} \biggl[j^{(1)} \delta \left(x-\dfrac{d}{2}\right) + j^{(2)} \delta
\left(x+\dfrac{d}{2}\right) \biggr] {\bf z}_0\:,
\end{aligned}
\end{equation}
where $j^{(1,2)}$ are the surface current densities induced in the graphene layers placed at $x= \pm d/2$, as indicated by Dirac's delta functions $\delta$. Assuming the field to be p-polarized with the magnetic component ${\bf H} = H(x,z,t) {\bf y}_0$, from Eq.~(\ref{eq:eqMaxw2}) we find
\begin{equation} \label{eq:eqMaxwH}
\frac{\varepsilon}{c^2} \frac{\partial^2 H}{\partial t^2} - \Delta H \! = \!  - \frac{4\pi }{c}  \frac{\partial }{\partial x} \left[ j^{(1)} \delta \left(x-\dfrac{d}{2}\right) + j^{(2)} \delta \left(x+\dfrac{d}{2}\right) \right]\:,
\end{equation}
where, if we assume the harmonic field dependencies $\sim \exp(-i \omega t)$, the currents are given by
\begin{equation}
 j^{(1,2)} (\omega, z) = \sigma^{(1,2)} (\omega, z) E_z (x = \pm {d}/{2}, z, \omega) \:.
\end{equation}
Here $\sigma^{(2)} (\omega, z) = \sigma (\omega ) $, $\sigma^{(1)} (\omega, z) = \sigma (\omega ) [1 + f(z)] $ are surface conductivities, $\sigma (\omega ) \equiv \sigma^{(R)} (\omega )  + i \sigma^{(I)}(\omega ) $ is the linear frequency-dependent surface conductivity of graphene, while $ f(z) = f(z + a) = f_1 \cos \left(\dfrac {2 \pi } {a} z \right)+ f_2 \cos \left(\dfrac {4 \pi } {a} z \right)$ is assumed to be a periodic function of $z$, i.e. conductivity of one of the graphene layers is spatially modulated.

Assuming the dissipative losses and modulation amplitudes $f_{1,2}$ to be small, we formally introduce the smallness parameter $\mu$
\begin{equation}
\mu = \text{max}  \left\{   \left| \dfrac{\sigma^{(R)}}{\sigma^{(I)}}\right|, f_{1,2} \right\}\:,
\end{equation}
and adopt the following asymptotic ansatz for the magnetic field
\begin{equation}
\begin{aligned}  \label{eq:ansatz}
H(x,z,t)& = e^{-i \omega t} \biggl \{ \mu \bar{H}_0 (x, \mu z)  \\
&  + \left[  \mathcal{ A }_{1}(\mu z) h(\omega, x) + \mu \tilde {H}_1 (x, \mu z) \right] e^{i k_{\text{sp}}(\omega) z}  \\
& + \left[  \mathcal{ A }_{2}(\mu z) h(\omega, x) + \mu \tilde {H}_2 (x, \mu z) \right] e^{- i k_{\text{sp}}(\omega) z} \biggr \}\:,
\end{aligned}
\end{equation}
where $h(x)$ is the transverse profile of the linear plasmonic mode of the guiding structure, $\mathcal{ A }_{1,2}$ are slowly varying mode amplitudes, where the subscripts $"1"$ and $"2"$ refer to the forward and backward propagation along the $z$ direction, respectively. Also $\tilde {H}_{1,2}$ are small corrections to the eigenmode profile, and the term $\bar{H}_0 $ does not contain a fast dependence on $z$.

In the zero-order approximation in $\mu$, % implying $H(x,z,t) =  h(x) e^{-i (\omega t - k_{\text{sp}} z)}$, $e_z(x) = \displaystyle{\frac{i}{k_0 \varepsilon} \frac{dh}{dx} }$,
for the function $h(x)$ Eq.~(\ref{eq:eqMaxwH}) takes the form,
\begin{equation}
\begin{aligned}
& \frac{d^2 h}{dx^2} -  (k_{\text{sp}}^2 - k_0^2 \varepsilon )h(x) = \\
& = \frac{4\pi }{c} i \sigma^{(I)} (\omega ) \frac{\partial }{\partial x}
\biggl\{ \left[{\delta} \left(x-\dfrac{d}{2}\right) + {\delta} \left(x+\dfrac{d}{2}\right) \right] e_z(x) \biggr\}
\:,
\end{aligned}
\end{equation}
where $e_z(x) = \displaystyle{\frac{i}{k_0 \varepsilon} \frac{dh}{dx} }$.
This equation yields the dispersion relation
\begin{equation}  \label{eq:DispRel1}
% 1 + \dfrac{2\pi i \sigma \kappa}{\omega \varepsilon}(1 \pm e^{-\kappa d}) = 0 \:,
\left[1 - \dfrac{2\pi \sigma^{(I)} \kappa^{{\text{(s,a)}}}}{\omega \varepsilon}(1 \pm e^{-\kappa^{\text{(s,a)}} d})\right] = 0 \:,
\end{equation}
where $\kappa^{\text{(s,a)}} = \sqrt{k^{\text{(s,a)} 2}_{\text{sp}} - k_0^2 \varepsilon}$, $k_0 = \omega / c$,
and the modal transverse profiles are
\begin{equation} 
h^{\text{(s,a)}}(\omega, x) = \frac{-ik_0 \varepsilon}{\kappa^{\text{(s,a)} 2}} \frac{de^{\text{(s,a)}}_z}{dx} \:,
\end{equation}
where the continuous tangential components of the electric field are expressed as
\begin{equation} \label{eq:TrStr1}
 e^{\text{(s)}}_z(\omega, x) =
	\begin{cases}
		e^{-\kappa^{\text{(s)}} (x-d/2)},&\text{}\ x>  d/2\:,\\
		 \displaystyle{\frac{\cosh(\kappa^{\text{(s)} }x)}{\cosh(\kappa^{\text{(s)}} d/2)}},&\text{}\ |x|<  d/2\:,\\
		e^{\kappa^{\text{(s)}} (x+d/2)},&\text{} x< - d/2\:
	\end{cases}
\end{equation}
and
\begin{equation} \label{eq:TrStr1}
  e^{\text{(a)}}_z(\omega,x) =
	\begin{cases}
		e^{-\kappa^{\text{(a)}} (x-d/2)},&\text{}\ x>  d/2\:,\\
		 \displaystyle{\frac{\sinh(\kappa^{\text{(a)}} x)}{\sinh(\kappa^{\text{(a)}} d/2)}},&\text{}\ |x|<  d/2\:,\\
		-e^{\kappa^{\text{(a)}} (x+d/2)},&\text{} x< - d/2\:
	\end{cases}
\end{equation}
for both symmetric and antisymmetric eigenmodes guided by a double-layer waveguide~\cite{buslaev_JETP,MultiWg}, as indicated by the superscripts.

We now assume that the period of the conductivity grating is chosen to almost compensate the momentum mismatch with the antisymmetric plasmon, $\displaystyle{|k_{\text{sp}}^{\text{(a)}}(\omega) - {2 \pi}/{a} | \equiv |\Delta k| \ll {2 \pi}/{a} }$ so that $|\Delta k /k_{\text{sp}}^{\text{(a)}}(\omega)|  \lesssim \mu $.
Substituting the expression of the total field~(\ref{eq:ansatz}) into Eq.~(\ref{eq:eqMaxwH}) and keeping only resonant terms, in the first order in $\mu$ for the field perturbations $\tilde{H}_{1,2}$, we obtain
\begin{equation} \label{eq:H2}
\begin{aligned}
& \dfrac{d^2 \tilde{H}_{1,2}}{dx^2} -  (k_{\text{sp}}^{{\text{(a)}}2} (\omega)- k_0^2 \varepsilon ) \tilde{H}_{1,2} - \dfrac{4 \pi}{c} i \sigma^{(I)} (\omega )  \\
& \times \frac{\partial }{\partial x}
\biggl\{ \left[{\delta}\left(x - \dfrac{d}{2}\right) + {\delta}\left(x + \dfrac{d}{2}\right) \right]  \tilde{E}_{1,2z} \biggr\} = F_{1,2}\:,\\
& F_{1,2} (x,z) = \mp 2ik_{\text{sp}}^{\text{(a)}}(\omega) \dfrac{\partial \mathcal A_{1,2}}{\partial z} h^{\text{(a)}}(\omega, x) + \frac{4 \pi}{c}  \sigma^{(R)} (\omega ) \mathcal A_{1,2} \\
&   \times  \frac{\partial }{\partial x}  \left\{ \left[ {\delta}\left(x - \dfrac{d}{2}\right) + {\delta} \left(x+\dfrac{d}{2}\right) \right]  e^{\text{(a)}}_z(\omega, x) \right\}  \\
&   + \frac{4 \pi}{c}  i \sigma^{(I)} (\omega )  \frac{\partial }{\partial x} \biggl\{  {\delta}\left(x - \dfrac{d}{2}\right) \left[ \frac{f_1} {2}\bar{E}_{0z} e^{\mp i\Delta k z} \right.  \biggr.   \\
& + \frac{f_2} {2}\mathcal A_{2,1}
\left. \biggl. e^{\mp 2i\Delta k z} e^{\text{(a)}}_z(\omega, x)  \right] \biggr\} \:,
\end{aligned}
\end{equation}
whereas $\bar{H}_{0}$ satisfies the equation,
\begin{equation} \label{eq:H0}
\begin{aligned}
\dfrac{d^2 \bar{H}_{0}}{dx^2} &\!+\! k_0^2 \varepsilon \bar{H}_{0} \! = \! \dfrac{4 \pi}{c} i \sigma^{(I)} (\omega ) \frac{\partial }{\partial x}
\biggl \{ \left [{\delta}\left(x + \dfrac{d}{2}\right) \!+ \!{\delta}\left(x - \dfrac{d}{2}\right) \right]\\
&  \times  {E}_{0z} (\omega,x)
\biggl. +\frac{f_1} {2} \left(\mathcal A_{1} e^{ i\Delta k z}  + \mathcal A_{2} e^{-i\Delta k z} \right) {\delta} \left(x-\dfrac{d}{2}\right)  \biggr \} \:.
\end{aligned}
\end{equation}
Equation~(\ref{eq:H0}) includes reflection of a plane wave incident from the semi-space $x>d/2$, written as $H_0{\rm exp} [-ik_{0}\sqrt{\varepsilon}(x-d/2)]$, where we assume $H_0 \sim \mu$, $E_0 = H_0/\sqrt{\varepsilon}$, ${E}_{0z} (\omega,x)$ is the tangential electric field distribution at $f_{1,2}=0$. Taking into account the boundary conditions at the graphene interface, we find the amplitude of the electric field at $x=d/2$ as follows
\begin{equation} \label{eq:E0}
\begin{aligned}
 & \bar{E}_{0z} (\omega, x=d/2) = E_0(1+r) \\
 & - \frac{2 \pi}{c \sqrt{\varepsilon}} i  \sigma^{(I)}(\omega )  \frac{f_1} {2} \left(\mathcal A_{1} e^{ i\Delta k z}  + \mathcal A_{2} e^{-i\Delta k z} \right) \:.
\end{aligned}
\end{equation}
where $r = - \displaystyle{{ \frac{4\pi} {c \sqrt{\varepsilon}} i \sigma^{(I)}(\omega ) }\left({1+\frac{4\pi} {c \sqrt{\varepsilon}}  i \sigma^{(I)} (\omega ) } \right)^{-1} } \approx - { \frac{4\pi} {c \sqrt{\varepsilon}} i \sigma^{(I)}(\omega ) }$
stands for the reflection coefficient from two closely spaced graphene layers, $k_0 \sqrt{\varepsilon} d  \ll 1$.% ~\cite{bludov_JO_2013}.

Substituting the expression~(\ref{eq:E0}) into Eq.~(\ref{eq:H2}),
in order for the corrections $\tilde{H}_{1,2}$ to be nondiverging, in accord with the Fredholm alternative~\cite{Korn},
we have to satisfy the orthogonality condition for the right-hand side of Eq.~(\ref{eq:H2}) with the solution of its homogeneous part, or equivalently, with the plasmonic mode itself.
This overintegrating is mathematically written as %
\begin{equation} 
\int\limits_{-\infty}^{+\infty}F_{1,2}(x,z){h^{\text{(a)}}}^{*}(\omega, x)dx = 0,
\end{equation}
and it leads to the nonlinear equations for the slowly varying amplitudes $\mathcal A_{1,2}$
\begin{multline} \label{eq:EqA}
\pm 2ik^{\text{(a)}}_{\text{sp}} (\omega) \left(\dfrac{\partial \mathcal A_{1,2}}{\partial z} \pm (\gamma^{\text{(a)}}  + \gamma^{\text{(a)}} _{r} ) \mathcal A_{1,2} \right)  {} \\
 {}\!=\!(\!  - 2ik^{\text{(a)}}_{\text{sp}} (\omega) \gamma^{\text{(a)}}_{r}\!+Q^{\text{(a)}} _2 ) \mathcal A_{2, 1} e^{ \mp 2 i\Delta k z} + Q^{\text{(a)}} _1 E_0 (1+r) e^{ \mp i\Delta k z} \:,
\end{multline}
where linear damping due to losses in graphene $\gamma^{\text{(a)}}$, grating-induced damping due to radiation $\gamma^{\text{(a)}}_r$, and coupling coefficients $Q^{\text{(a)}} _{1,2} $ are derived, respectively, in the form
\begin{gather}
\gamma^{\text{(a)}}  =  \dfrac{4\pi}{c}  \sigma^{(R)} (\omega ) \dfrac{k_0 \varepsilon}{k^{\text{(a)}}_{\text{sp}}(\omega) q^{\text{(a)}} } \:,\nonumber \\
\gamma^{\text{(a)}}_r =  \left(\dfrac{\pi \sigma^{(I)}(\omega ) f_{1} }{ c} \right)^2 \dfrac{k_0 \varepsilon^{1/2}}{k^{\text{(a)}}_{\text{sp}} (\omega) q^{\text{(a)}} } \:, \nonumber \\
Q^{\text{(a)}}_{1,2} = \dfrac{2\pi}{c} \sigma^{(I)} (\omega )  f_{1,2} \dfrac{k_0 \varepsilon}{ q^{\text{(a)}}} \:,
\end{gather}
where in all the expressions we employ the following definition,
\begin{equation}
q^{\text{(a)}} = {k_0^2} \dfrac{\varepsilon^2}{\kappa^{\text{(a)}3} } \left[ 1 + \dfrac{\sinh{(\kappa^{\text{(a)}} d)} + \kappa^{\text{(a)}} d }{2 \sinh^2 {(\kappa^{\text{(a)}} d /2)}} \right].
\end{equation}

When we assume a simple homogeneous case, $\dfrac{\partial }{\partial z} =0$,
and also the exact momentum matching, $\Delta k=0$,  from Eq.~(\ref{eq:EqA}) we obtain
the slowly varying amplitudes,
\begin{equation}
\mathcal A_{1} = \mathcal A_{2} \equiv \mathcal A = \dfrac{Q^{\text{(a)}}_{1}  E_0 (1+r)}{2ik^{\text{(a)}}_{\text{sp}}  (\omega) [\gamma^{\text{(a)}} (\omega) + 2 \gamma^{\text{(a)}}_r] - Q^{\text{(a)}}_{2}}
\end{equation}
and substitute them into Eq.~(\ref{eq:E0}). Thus, the excitation of the antisymmetric mode results in changing the reflection coefficient compared to the case without a conductivity grating,
$r_{\mathcal A} \approx r(1 + f_1 \mathcal A /2 E_0)$.

\section{Nonlinear plasmon-to-plasmon conversion}

First we notice that, for the case of TM-polarized waves with the tangential electric field of the form $\mathbf{E}_{\tau \omega} = \mathbf{z}_0 E e^{iqz}$ (the monochromatic time dependence $\sim \exp (-i \omega t)$ is omitted),
the induced second-harmonic current in graphene depends on the in-plane momentum
$q$ manifesting its nonlocal nature. It is expressed as~\cite{Mikhailov2011_SHG, SHG_Wrapped}
\begin{equation}
{\bf j}_{2 \omega} =  - \mathbf{z}_0 \dfrac{3} {8} \dfrac{e^3 v_{F}^2} {\pi \hbar^2 \omega^3} q E^2 e^{i2qz} \:,
\end{equation}
where $v_{F}\approx c/300$ is the Fermi velocity. This result is derived in the semiclassical limit from the Boltzmann kinetic equation written for Dirac electrons with a linear energy dispersion under the approximations $\hbar \omega \leq \mathcal{E}_{F}$, $ k_B T \ll \mathcal{E}_{F}$ and $q v_{F}/ \omega \ll 1$, where $\mathcal{E}_{F}$ is the Fermi energy, $k_B$ is the Boltzmann constant, $T$ is the temperature. Thereby, a nonlinear source associated with the antisymmetric mode will generate a synchronous symmetric mode. Importantly, in general, when discussing the parametric
interaction that involves the modes of different symmetries, the phase matching is not a sufficient condition, and the overlaps of the modes with the respective nonlinear sources should be examined. Since a quadratic nonlinear source localized at the graphene layers is symmetric, the conversion between the FF antisymmetric mode and SH symmetric mode becomes possible.

Similar to the analysis of the fundamental frequency field, we employ here the perturbational method
 and derive the equations for the envelopes of the second-harmonic fields $\mathcal B_{1, 2}$,
\begin{multline} \label{eq:EqB}
\pm 2ik^{\text{(s)}}_{\text{sp}} (2\omega)  \left(\dfrac{\partial \mathcal B_{1,2}}{\partial z} \pm (\gamma^{\text{(s)}}  + \gamma^{\text{(s)}} _{r} ) \mathcal B_{1,2} \right)  {} \\
 {} =  - 2ik^{\text{(s)}}_{\text{sp}} (2\omega) \gamma^{\text{(s)}}_{r}  \mathcal B_{2, 1} e^{ \mp 2 i (\widetilde{\mathtt{\Delta} k } + 2 \Delta k) z} + g \mathcal A_{1, 2}^2 e^{ \mp  i \widetilde{\mathtt{\Delta} k } z}    \:,
\end{multline}
where $\widetilde{ \mathtt{\Delta} {k} }= k^{\text{(s)}}_{\text{sp}}(2\omega) - 2 k^{\text{(a)}}_{\text{sp}}(\omega)$ is a small detuning, and other parameters are
\begin{gather}
q^{\text{(s)}} = 4 {k_0^2} \dfrac{\varepsilon^2}{\kappa^{\text{(s)}3}} \left[ 1 + \dfrac{ \sinh{(\kappa^{\text{(s)}} d)} - \kappa^{\text{(s)}} d }{2 \cosh^2 {(\kappa^{\text{(s)}} d /2)}} \right]\:, \nonumber   \\
\gamma^{\text{(s)}}  =  \dfrac{8\pi}{c}  \sigma^{(R)} (2 \omega ) \dfrac{k_0 \varepsilon}{k^{\text{(s)}}_{\text{sp}} (2\omega) q^{\text{(s)}} }\:,\nonumber \\
\gamma^{\text{(s)}}_r =  \left(\dfrac{\pi \sigma^{(I)}(2 \omega ) f_{2} }{ c} \right)^2 \dfrac{2 k_0 \varepsilon^{1/2}}{k^{\text{(s)}}_{\text{sp}} (2\omega) q^{\text{(s)}} }\:,\nonumber \\
g = -i \dfrac{16\pi}{c} \sigma_{2}(\omega) \dfrac{k_0 \varepsilon}{ q^{\text{(s)}} }\:,
\end{gather}
with  $\sigma_{2} (\omega)$ denoting the quadratic conductivity of graphene defined as
\[
\sigma_{2} (\omega) = -  \dfrac{3} {8} \dfrac{e^3 v_{F}^2} {\pi \hbar^2 \omega^3} k^{\text{(a)}}_{\text{sp}} (\omega).
\]
Within the undepleted pump approximation, when the
amplitude at fundamental frequency is not affected by nonlinearity, for the exact phase matching, $\widetilde{\mathtt{\Delta} k }= \Delta k =0$, and homogeneous case discussed above, we have
\begin{equation}
\mathcal B_{1} = \mathcal B_{2} \equiv \mathcal B = \dfrac{g \mathcal A^2}{2ik^{\text{(s)}}_{\text{sp}} (2\omega) (\gamma^{\text{(s)}} + 2 \gamma^{\text{(s)}}_r) }.
\end{equation}

Since the radiation associated with SH field is emitted equally to both directions, the normalized conversion efficiency can be introduced as a ratio of the doubled SH energy flux density in the upper half-space to the energy flux of the incident fundamental wave. Equivalently, this definition takes the following form
\begin{equation}
\eta = \dfrac{2|\bar{E}_{0z} (2\omega, x=d/2)|^2}{|{E}_{0}|^2},
\end{equation}
where $\bar{E}_{0z} (2\omega, x=d/2) = - \dfrac{2 \pi}{c \sqrt{\varepsilon}} i  \sigma^{(I)}(2\omega )  f_2 \mathcal B $.

% ------------------------------------------------------------------------

\section{Discussions}
\begin{figure}[t]
\centering\includegraphics[width=0.95\linewidth]{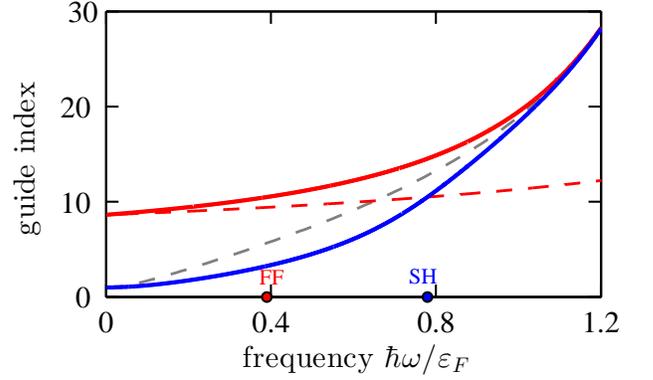}
\caption{(Color online) Plasmonic guide index vs. frequency at the fixed separation between the graphene layers, $d=123$ nm, calculated for the doping level of $\mathcal{E}_{F}=0.3$ eV, $N = 5$, $\varepsilon = 1$.
Blue and red solid curves correspond to the symmetric and antisymmetric modes, respectively.
Dashed gray and red lines show dispersion of an isolated graphene sheet and half-frequency antisymmetric mode, for reference.
Color dots in the horizontal axis mark the fundamental and second-harmonic frequencies.}
\label{fig:PhMatch}
\end{figure}
\begin{figure}[t]
\centering\includegraphics[width=0.95\linewidth]{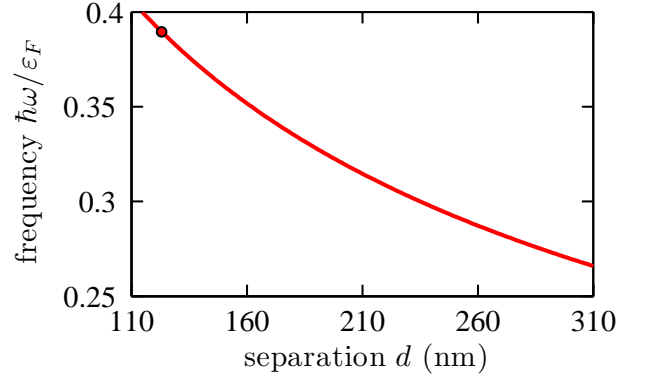}
\caption{(Color online) Operational fundamental frequency as a function of the separation between the layers for the exact phase-matching at $\mathcal{E}_{F}=0.3$ eV, $N = 5$, $\varepsilon = 1$. The point corresponds to the parameters used in Fig.~\ref{fig:PhMatch}.
}
\label{fig:dDep}
\end{figure}

In the framework of the nonlinear amplitude equations, we can now analyze the parametric effects and
estimate the frequency conversion efficiency. To make a period of the conductivity modulation realistic, in our calculations we employ multilayer graphene that effectively increases the equivalent surface conductivity leading to a reduction of the wavenumber of the p-polarized plasmons supported by multilayer graphene structures~\cite{DissipSoliton_LPR, MultiWg}. We take ${\sigma(\omega)}= N {\sigma_s}(\omega)$ for a randomly stacked multilayer graphene film consisting of $N$ layers, with each layer characterized by a surface conductivity~\cite{sptdep}
\begin{equation}
\sigma_{s}(\omega) = \displaystyle{\frac{ie^2}{\pi \hbar}
\left\{\frac{\mathcal{E}_{F}}{\hbar\left(\omega + i\tau_{\text{intra}}^{-1}\right)} + \frac{1}{4} \text{ln}\left|\frac{2\mathcal{E}_{F} -\hbar \omega}{2\mathcal{E}_{F} + \hbar \omega}\right|\right\}}\:,  \label{eq:sigma}
\end{equation}
where for doped graphene we assume $\hbar\omega < 1.67\mathcal{E}_{F} $ and $k_B T \ll \mathcal{E}_{F}$, and $\displaystyle{\tau_{\text{intra}} ^{-1}}$ is the relaxation rate.

To reveal a possibility of plasmonic frequency conversion in the graphene waveguide, in Fig.~\ref{fig:PhMatch} we plot the frequency-dependence of the plasmonic guide indices for the symmetric and antisymmetric modes.
We observe that the first harmonic antisymmetric and second-harmonic symmetric modes of the waveguide can be matched at $\hbar \omega = 0.12$ eV ($\lambda=2\pi/k_0\approx 10.6$ $\mu$m, corresponding to one of the radiation lines of high power $\text{CO}_2$ lasers), $ k^{\text{(a)}}_{\text{sp}}  \approx 10.5 k_0$.

Figure~\ref{fig:dDep} shows the dependence of the fundamental frequency on the separation between the layers.
Note while discussing the dispersion relations and the possibility of
phase-matching conditions we take $\sigma = i  \sigma^{(I)} $ and do not take losses into account,
however these losses are included into our considerations perturbatively. Calculated for $f_1=f_2=0.1$, $a \approx 1 $ $\mu$m, $\tau_{\text{intra}} = 0.3$ ps, the incident wave intensity of % 100 MW/cm^2 $10^{8}$ W/cm$^2$
$1$ MW/cm$^2$  and the other parameters of Fig.~\ref{fig:PhMatch}, efficiency is $\eta \sim 8.7 \times 10^{-7}$. Overall, for this structure the smallness parameter is found to be $\mu \sim 0.1$, so that our asymptotic approach is well justified.

\section{Conclusions}

By employing the approximation of slowly-varying field amplitudes, we have derived nonlinear equations for describing the second-harmonic generation from a double-layer graphene structure with modulated conductivity. We have revealed
that the quadratic phase matching between the plasmonic modes of different symmetries becomes possible in a planar waveguide geometry with conductivity modulation playing a role of an effective grating that couples the external pumping radiation to the waveguiding modes with the subsequent parametric conversion into radiation at the double frequency.

\section*{ACKNOWLEDGMENTS}

The authors are grateful to I.V. Shadrivov and A.I. Smirnov for in-depth comments,
and also thank S.A. Mikhailov for useful discussions at an early stage of this project.
This work was supported by the Australian National University.

% \bibliography{my}

\end{document}